# The role of low-energy phonons with mean-free-paths >0.8 µm in heat conduction in silicon

**Puqing Jiang and Yee Kan Koh[*]**

Department of Mechanical Engineering, National University of Singapore, Singapore 117576

[*]corresponding author: mpekyk@nus.edu.sg

## ABSTRACT

Despite recent progress in the first-principles calculations and measurements of phonon mean-free-paths ($\ell$), contribution of low-energy phonons to heat conduction in silicon is still inconclusive, as exemplified by the discrepancies as large as 30% between different first-principles calculations. Here we investigate the contribution of low-energy phonons with $\ell$>0.8 µm by accurately measuring the cross-plane thermal conductivity ($\Lambda_{\text{cross}}$) of crystalline silicon films by time-domain thermoreflectance (TDTR), over a wide range of film thicknesses $1 \leq h_f \leq 10$ µm and temperatures $100 \leq T \leq 300$ K. We employ a dual-frequency TDTR approach to improve the accuracy of our $\Lambda_{\text{cross}}$ measurements. We find from our $\Lambda_{\text{cross}}$ measurements that phonons with $\ell$>0.8 µm contribute 53 W m$^{-1}$ K$^{-1}$ (37%) to heat conduction in natural Si at 300 K while phonons with $\ell$>3 µm contribute 523 W m$^{-1}$ K$^{-1}$ (61%) at 100 K, >20% lower than the first-principles predictions by Lindsay *et al*. of 68 W m$^{-1}$ K$^{-1}$ (47%) and 695 W m$^{-1}$ K$^{-1}$ (77%), respectively. Using a relaxation times approximation (RTA) model, we demonstrate that macroscopic damping (e.g., Akhieser's damping) eliminates the contribution of phonons with mean-free-paths >30 µm at 300 K, which contributes 15

W m$^{-1}$ K$^{-1}$ (10%) to heat conduction in Si according to Lindsay *et al*. Thus, we propose that omission of the macroscopic damping for low-energy phonons in the first-principles calculations could be one of the possible explanations for the observed differences between our measurements and calculations by Lindsay *et al*. Our work provides an important benchmark for future measurements and calculations of the distribution of phonon mean-free-paths in crystalline silicon.

## Text

Phonons are the dominant energy carriers in most semiconductors and dielectric materials.[1] For many decades, understanding of heat transport by phonons is incomplete, mainly due to lack of quantitative knowledge[2] of how far phonons propagate without being scattered, a property called the mean-free-paths ($\ell$) of phonons. Recently, significant breakthroughs are attained with successful calculations of phonon mean-free-paths without any fitting parameters using first-principles approaches,[3-5] and good agreement was achieved between the first-principles calculations and experimental data of bulk thermal conductivity for a wide range of materials.[6-9] This exciting development leads to predictions of the thermal conductivity of new materials[10] that are yet to be discovered.

Despite these impressive achievements, disagreement still exists between first-principles calculations by different researchers, especially for low-energy phonons with long mean-free-paths, see Fig. S7 in the online supporting materials for a comparison of published first principles calculations of isotopically pure Si at room temperature. For example, Lindsay *et al.*[11, 12] calculated that phonons with mean-free-paths $\ell > 0.8$ μm contribute ≈70 W m$^{-1}$ K$^{-1}$ (47%) to thermal transport in isotopically pure Si, but Garg *et al.*[13] estimated a contribution of only ≈40 W m$^{-1}$ K$^{-1}$ (30%), even though the contributions of high-energy phonons calculated by both groups are similar (≈90 W m$^{-1}$ K$^{-1}$). While this paper was under review for publication, Jain and McGaughey published their first-principles calculations of isotopically pure Si using different exchange-correlation and pseudopotential schemes to represent the many-body interactions of electrons within the framework of density function theory (DFT).[14] They found that the calculated thermal conductivities vary by 30% when different DFT schemes were employed, demonstrating that the accuracy of first-principles calculations hinges on a close approximation of the computationally expensive many-body interactions of electrons. Moreover, in the first-principles approaches, researchers do not take into account attenuation of low-energy phonons due to macroscopic mechanisms,[15] such as irreversible viscous absorption (often called Akhieser's damping) and thermoelastic energy loss, which are proven to dominate scattering of low-energy phonons.[16-18] In fact, Maznev estimated that the onset of measurable size effects of Si is reduced from ≈100 μm as predicted by first-principles calculations to ≈10 μm if the macroscopic mechanisms are taken into account.[19]

In principle, the contribution of low-energy phonons can be verified by measurements of micron-sized Si structures. Goodson and his coworkers[20, 21] measured the in-plane thermal conductivity of Si thin films with film thickness $h_f$<1 μm by the 3ω method, and, contrary to the first-principles calculations, they observed no significant reduction in the thermal conductivity in their 1-μm thick film. More recently, Cuffe et al.[22] reported in-plane thermal conductivity of Si membranes of 15 nm–1.5 μm using transient thermal grating method, and they reconstructed a thermal conductivity accumulation function that compares fairly well with the first-principles calculations by Esfarjani et al.[4] In these experiments, however, the focus was on the contribution of higher-energy phonons, considering the thinner films with $h_f$<1 μm studied in the experiments and the fact that boundary scattering of phonons is weak in the in-plane direction[23] when $h_f \ll \ell$.

Meanwhile, researchers also attempted to directly probe the phonon mean-free-paths by imposing a temperature profile with a characteristic length $L_c$ comparable to the mean-free-paths of phonons, by changing either the frequency[24, 25] or the size of the heat source[26-28] in the experiments. In these mean-free-path spectroscopy techniques, measurements are usually interpreted[24-26] using an empirical assumption first proposed by Koh et al.[24] that phonons with $\ell > L_c$ are ballistic and do not contribute to heat conduction. Such simple interpretation, however, is not always valid because factors such as the anisotropy[29] in heat transport by the non-equilibrium phonons, the effects of interfaces[29] and the diminished relative phase[30] between the applied heat flux and the temperature response at high frequencies cannot always be ignored. The challenges in the analysis and interpretation of measurements using the mean-free-path spectroscopy techniques lead to conflicting conclusions on the contribution of phonons with $\ell > 0.8$ µm in Si.[24, 25, 29]

In this paper, we examine the reduction of the thermal conductivity in silicon thin films in the cross-plane direction to determine the contribution of low-energy phonons to heat conduction in Si. We accurately measured the cross-plane thermal conductivity ($\Lambda_{cross}$) of crystalline Si films with thickness $1 \leq h_f \leq 10$ µm at temperatures $100 \leq T \leq 300$ K, using a dual-frequency time-domain thermoreflectance (TDTR) approach. While our $\Lambda_{cross}$ measurements fall between the wide range of predictions derived from the first-principles calculations, our $\Lambda_{cross}$ measurements are higher than widely cited[22, 25, 26, 31] calculations by Lindsay et al.[11] and Esfarjani et al.[4] Specifically, at 300 K, we measured $\Lambda_{cross}=90$ W m$^{-1}$ K$^{-1}$ for $h_f=1$ µm, compared to first principles predictions of 78 W m$^{-1}$ K$^{-1}$.[4,11] The difference is beyond the uncertainty of our measurements. We build a relaxation time approximation (RTA) model to evaluate whether the difference could be attributed to the macroscopic damping not considered in the first-principles calculations. We find that for Si, the macroscopic damping strongly attenuates phonons with $\ell>30$ µm, 200 µm and 600 µm at $T$=300 K, 150 K and 100 K. The reduction of thermal conductivity due to macroscopic damping is on the same order of magnitude as the difference between our measurements and Lindsay's calculations.

**Experimental details**

Our samples are commercially available p-type <100> silicon-on-insulator (SOI) wafers with a device layer of 1−10 µm. The 1 µm thick SOI wafer was prepared by the Smart Cut$^{TM}$ process while other SOIs were prepared by the bonding and etch-back process.[32] The device layers are single-crystalline, with a resistivity of 10−20 Ω-cm (measured by a four-point probe) and an estimated impurity concentration of $\approx 10^{15}$ cm$^{-3}$. With this level of impurity concentration, the strength of Rayleigh scattering[33] due to the impurity is four orders of magnitude smaller than that due to natural isotopes of Si; thus the effect of impurity is negligible. To prepare the samples for TDTR measurements, we first etched away the native silicon oxide of the SOIs using hydrofluoric acid (HF) and then immediately deposited a $\approx$100 nm thick Al film. We measured the thicknesses of Al, Si and SiO$_2$ layers by picosecond acoustics.[34] Since accurate thickness of the Al transducer with native oxide is important, we measured the thickness of the native Al oxide layer as 3.7±0.5 nm from the Al 2p X-ray photoelectron spectroscopy (XPS) spectrum.[35, 36]

We employed time-domain thermoreflectance (TDTR)[37] to measure the cross-plane ($\Lambda_{cross}$) thermal conductivities of the Si films. In TDTR measurements, a train of femtosecond laser pulses is split into a pump beam and a probe beam. The pump beam, modulated by an electro-optic modulator, is absorbed by the transducer of the sample and periodically heats the sample at a modulation frequency $f$. The periodic temperature response at $f$ at the sample surface is then monitored by a synchronized but time-delayed probe beam via thermoreflectance, using a photodiode detector and a lock-in amplifier. With the periodic heating at the sample surface, heat diffuses a distance $d_p$ into the sample, where $d_p = \sqrt{\Lambda/\pi C f}$ is the thermal penetration depth[24] and $C$ is the volumetric heat capacity. Thus, the temperature response is only sensitive to thermal properties within the distance $d_p$ of the sample. We then extract the thermal conductivity of the Si films by comparing the ratio of in-phase and out-of-phase signals of the lock-in amplifier at $f$, $R_f = -V_{\text{in}} / V_{\text{out}}$, to calculations of a diffusive thermal model.[38] We note that we used an anisotropic thermal model in the analysis of our TDTR measurements, although this is not critical because we designed the experiments such that the measurements are sensitive to $\Lambda_{\text{cross}}$ but not $\Lambda_{\text{in}}$, see below.

In order for TDTR to have high enough sensitivity to $\Lambda_{\text{cross}}$, we applied a large laser spot size (i.e., $w_0 = 27$ µm) and a high modulation frequency $f_h$ to ensure that $d_p \ll w_0$ and thus one-dimensional heat transfer in the cross-plane direction; in this case, TDTR measurements are sensitive to $\Lambda_{\text{cross}}$ not $\Lambda_{\text{in}}$. However, accurate measurements of $\Lambda_{\text{cross}}$ are still challenging. The TDTR measurements at $f_h$ are usually far more sensitive to the thickness of the Al transducer $h_{\text{Al}}$ due to the low thermal resistance of the Si films, yielding an unacceptably large uncertainty for the $\Lambda_{\text{cross}}$ measurements.

Here, we develop a dual-frequency TDTR approach to improve the accuracy of our $\Lambda_{\text{cross}}$ measurements. We observe that for TDTR measurements on similar geometries, the sensitivity of TDTR signals to $h_{\text{Al}}$ depends only weakly on the modulation frequency $f$, while the sensitivity to $\Lambda_{\text{cross}}$ reduces drastically as $f$ decreases. Thus, we perform an additional TDTR measurement at a lower modulation frequency $f_0$, such that the sensitivity to $h_{Al}$ is comparable to that at $f_h$ and the sensitivity to $\Lambda_{cross}$ is near zero. We then derive $\Lambda_{cross}$ of the Si thin films by adjusting calculations of the thermal model to fit the ratio of TDTR signals at the two frequencies $f_h$ and $f_0$, $R_{dual} = R_{f_h} / R_{f_0}$, see Fig. 1(a). By analyzing $R_{dual}$, the sensitivity to $\Lambda_{cross}$ is maintained, while the sensitivity to $h_{Al}$ is greatly reduced; the uncertainty of our $\Lambda_{cross}$ measurements is thus improved to $\approx 10\%$. The advantage of the dual-frequency approach is demonstrated in Fig. 1(b), in which we compare calculations of the thermal model using the $\Lambda_{cross}$ fitted from the $R_{dual}$ to the TDTR measurements at each individual frequency $f_h$ and $f_0$. As shown in Fig. 1(b), the agreement, while acceptable, is not perfect, due to additional errors from the parameters (e.g., $h_{Al}$) that only $R_f$ (but not $R_{dual}$) is sensitive to. More details on this dual-frequency approach, including the choice of $f_h$ and $f_0$, can be found in the online supporting material and Ref. [39].

Our measurements are not affected by the frequency dependence[24, 25] and laser spot size dependence[26] observed in prior pump-probe thermoreflectance measurements. We verify that frequency dependence is negligible for bulk Si when the laser spot size is sufficiently large (i.e., for $w_0$=27 µm used in our $\Lambda_{cross}$ measurements), see Fig. 2, consistent with several prior TDTR measurements on Si using large laser spot sizes.[24, 29, 40] Slight frequency dependence is observed for our measurements at 100 K when a small laser spot size of $w_0$=5.5 µm was used, see Fig. 2. At 100 K, the penetration depths $d_p$ are 5 µm and 16 µm for modulation frequencies of 10 MHz and 1 MHz, respectively. Thus, the frequency dependence is a result of transition from measurements limited by a high modulation frequency ($d_p$<2$w_0$) to measurements limited by a small laser spot size ($d_p$>2$w_0$). We note that all subsequent $\Lambda_{cross}$ measurements reported in the manuscript were acquired using $w_0$=27 µm, and thus are not affected by the frequency dependence. Our $\Lambda_{cross}$ measurements are also not affected by spot size dependence because we use laser spot size much larger than the film thicknesses ($2w_0 > h_f$). In this case, the mean-free-paths of low-energy phonons are limited by $h_f$ due to the boundary scattering at the thin film interfaces and thus the nonequilibrium effects should be minimal.

## Results and discussions

We present our $\Lambda_{cross}$ measurements normalized by the thermal conductivity of bulk Si $\Lambda_{bulk}$ at respective temperatures[41] in Fig. 3 (a-c), and compare our $\Lambda_{cross}$ measurements with the thermal conductivity of Si nanowires,[42-44] and the apparent thermal conductivity measured on bulk Si using the broadband frequency-domain thermoreflectance (BB-FDTR)[25] and TDTR with different spot sizes.[26, 45] To facilitate the comparison, we plot the measurements as a function of a characteristic length $L_c$ responsible for the reduction in the measured thermal conductivity; see the caption of Fig. 3 for the definitions of $L_c$ for different experiments. Although these experiments are based on different underlying physics, the reduction in the measured thermal conductivity in all these experiments can be approximated with an additional scattering length of $L_c$; thus the comparison is justified. We note that the effective boundary scattering length of $3h_f/4$ for the $\Lambda_{cross}$ is derived from the radiation limit[46] for heat conduction in the cross-plane direction, assuming that heat is diffusely radiated and absorbed at the interfaces. Our $\Lambda_{cross}$ measurements compare favorably with prior measurements of thermal conductivity of Si nanowires, the spot size dependent TDTR measurements, and BB-FDTR measurements at room temperature, but disagree with BB-FDTR measurements at 150 K, see Fig. 3. The source of this disagreement is unknown to the authors.

We also compare our measurements to the first-principles calculations at 300 K in Fig. 3 (d). To make a fair comparison, we transform the first-principles calculations of the cumulative thermal conductivity of bulk Si into the corresponding thin film thermal conductivities, using an approach developed by Yang and Dames, see Eq. (12) of Ref. [31]. We use Matthiessen's rule to relate the phonon mean-free-paths in Si thin films $\ell_f$ and bulk Si $\ell_{bulk}$ as $\ell_f^{-1} = \ell_{bulk}^{-1} + L_c^{-1}$. We plot the $\Lambda_{cross}$ derived from the first-principles calculations of natural Si by Lindsay et al.,[11,12] and of isotopically pure Si by Li et al.,[47] Esfarjani et al.,[4] Garg et al.[13] and by Jain and McGaughey,[14] normalized by the respective $\Lambda_{bulk}$ calculated by the same researchers. We find that our measurements fall within the wide range of predictions by the first-principles approaches, see Fig. 3(d).

To quantify and compare the contribution of low-energy phonons to heat conduction in bulk Si, we define a cumulative thermal conductivity for low-energy phonons with mean-free-path longer than $\ell$ as $\Lambda_{low} = \int_{\ell}^{\infty} \Lambda_{\ell_{bulk}} d\ell_{bulk}$, where $\Lambda_{\ell_{bulk}}$ is the thermal conductivity contribution per $\ell_{bulk}$[31] and $\Lambda_{\ell_{bulk}} d\ell_{bulk}$ is the differential thermal conductivity in bulk Si by phonons with mean-free-path from $\ell_{bulk}$ to $\ell_{bulk} + d\ell_{bulk}$. Note that our definition of cumulative thermal conductivity $\Lambda_{low}$ is different from the typical definition of cumulative thermal conductivity which sums up the contribution of high-energy phonons with mean-free-path up to $\ell$. We then approximate $\Lambda_{low}$ for $\ell = L_c$ from our $\Lambda_{cross}$ measurements, $\Lambda_{low} = \Lambda_{bulk} - \Lambda_{cross}$, and plot $\Lambda_{low}$ as a function of phonon mean-free-path $\ell$ in Fig. 4. We choose this simple approximation instead of more rigorous reconstruction using a complex optimization procedure[48] because this convenient approach is sufficiently accurate within the range of our experiments, considering the uncertainty of our measurements; see the online supporting materials for full justification. We note that for the range of our experiments, this simple approximation is an overestimation and thus the experimental data in Fig. 4 represent an upper limit of $\Lambda_{low}$.

For comparison, we also plot the first-principles calculations of $\Lambda_{low}$ of natural Si by Lindsay *et al.* as solid lines in Fig. 4. While the calculations by Esfarjani *et al.*[4] have been widely cited for interpretations of prior experimental results,[22, 25, 26, 49] we choose to compare our measurements to calculations by Lindsay *et al.*[11] because i) Lindsay's approach, which includes the calculation of all the cubic force constants up to seventh nearest neighbors, was considered to be very accurate;[4, 50] ii) Lindsay's calculations are similar to those by Esfarjani *et al.*[4], see Fig. 3(d), yet Esfarjani's data are not available for the long mean-free-path range and need extrapolation, but Lindsay's data are available for the full spectrum; iii) Lindsay's calculations are for natural Si, which enables direct comparisons with our measurements, while all the other calculations are for isotopically pure Si.

We find from our measurements that phonons with $\ell$>0.8 µm contribute 53 W m$^{-1}$ K$^{-1}$ (37%) to heat conduction in Si at 300 K, while phonons with $\ell$>3 µm contribute to 523 W m$^{-1}$ K$^{-1}$ (61%) at 100 K, see Fig. 4. These values of $\Lambda_{low}$ are lower than the first-principles calculations by Lindsay *et al.* of 68 W m$^{-1}$ K$^{-1}$ (47%) and 695 W m$^{-1}$ K$^{-1}$ (77%), respectively. The differences of >20% are beyond the experimental uncertainty especially at cryogenic temperatures, see Fig. 4.

One possible explanation for the observed differences is that in our measurements some low-energy phonons are not scattered by the interfaces but are transmitted[51] across the interfaces. However, we argue that all transmitted phonons, if any, should be scattered by the underlying amorphous $SiO_2$ and thus the effective boundary scattering should be limited by $h_f$. To represent the uncertainty of the characteristic length for $\Lambda_{cross}$, we include error bars of $3h_f/4 - h_f$ in Figs. 3 and 4. We find that the disagreement persists even after the uncertainty is taken into consideration, see Figs. 3 and 4.

Another possible explanation for the discrepancy is that Lindsay *et al*. did not consider macroscopic damping that dominantly scatters low-energy phonons in their calculations, and thus overestimated the contribution of the low-energy phonons. There are at least two macroscopic mechanisms[15, 52] that could attenuate the low-energy phonons (i.e., ultrasonic waves). First, low-energy phonons create a periodic temperature profile of hotter (compressed) and colder (expanded) regions in Si, and heat transfer between these regions attenuates the low-energy phonons due to thermoelastic loss of energy. Second, strains generated by the low-energy phonons causes a change of the effective temperature of individual high-energy phonons (usually called the thermal phonons). As the thermal phonons relax to a new local equilibrium through e.g., the three-phonon processes, entropy is generated and the low-energy phonons are damped. This viscous damping is usually called the Akhieser's damping.[15]

To estimate the effect of macroscopic damping, we build a relaxation time approximation (RTA) model that successfully reproduces the accumulation functions by Lindsay *et al*. see Fig. S4 of the online supporting materials and Fig. 4 of the main text. Details of our RTA model are described in online supporting materials. In our RTA model, we calculate the phonon dispersion using an adiabatic bond-charge model,[53] and use the averaged values of the zeroth-order relaxation times from the first-principles calculations by Lindsay *et al.*[11, 12] To estimate the effects of the macroscopic damping to heat conduction in Si, we follow Maznev[19] to approximate the relaxation times due to the macroscopic damping using an expression $\tau_A = \tau_{inf}\left(1 + 1/\tau_{th}^2 \omega^2\right)$, where $\omega$ is the phonon angular frequency and $\tau_{inf}$ and $\tau_{th}$ are parameters derived from fitting of prior measurements of the relaxation times of ultrasonic waves[16-18] at different temperatures. Since measurements of the relaxation times of ultrasonic waves are available only up to $\omega$=100 GHz, this expression of $\tau_A$ is a simple approximation that smoothly fits both the experimental data for $\omega$<100 GHz and the first-principles calculations of phonon relaxation times due to three-phonon processes for $\omega$>1 THz. We incorporate the relaxation times $\tau_A$ into our RTA model using Matthiessen's rule. We calculate the cumulative thermal conductivity with and without the macroscopic damping using our RTA model. We then derive $\Lambda_{cross}$ using the approach by Yang and Dames, and plot the derived $\Lambda/\Lambda_{bulk}$ in Fig. 3. We also plot the $\Lambda_{low}$ calculated from our RTA model in Fig. 4.

We find that inclusion of the macroscopic damping essentially eliminates contribution of phonons with $\ell$>30 µm, 200 µm and 600 µm to heat conduction in Si at $T$=300 K, 150 K and 100 K, respectively, see Fig. 4. Since Lindsay *et al.* predicts a measurable contribution of 15 W m$^{-1}$ K$^{-1}$ (10%), 40 W m$^{-1}$ K$^{-1}$ (10%), and 120 W m$^{-1}$ K$^{-1}$ (15%), respectively, from these low-energy phonons, inclusion of the macroscopic damping reduces the cumulative thermal conductivity of low-energy phonons $\Lambda_{low}$ calculated by our RTA model to be in better agreement with our measurements, see Fig. 4. Thus, we propose that omission of macroscopic damping could be one of the possible explanations for the differences in the contribution of low-energy phonons between our measurements and the first-principles calculations by Lindsay *et al*.

## Conclusions

In summary, we find from our measurements of cross-plane thermal conductivity of Si films that phonons with $\ell>0.8$ µm contribute 37% to heat conduction in natural Si at 300 K and phonons with $\ell>3$ µm contribute 61% at 100 K, lower than 47% at 300 K and 77% at 100 K predicted by the first-principles calculations of Lindsay *et al*. We find that inclusion of the macroscopic damping in our RTA model eliminates the contribution of phonons with $\ell>30$ µm, 200 µm and 600 µm to heat conduction in Si at $T$=300 K, 150 K and 100 K, respectively, and thus could explain the differences between our measurements and the first-principles calculations. Our measurements also disagree with the BB-FDTR measurements at 150 K. Our measurements thus provide a crucial set of experimental data for comparison in future studies of the mean-free-paths of phonons in Si.

## Acknowledgements

We are grateful to Dr. L. Lindsay and Prof D. A. Broido for providing us their first-principles results of the cumulative thermal conductivity of both natural and isotopically pure silicon. We also sincerely thank Dr. L. Lindsay, R. B. Wilson and Prof D. G. Cahill for many useful discussions. This material is based upon work supported by the Singapore Ministry of Education, Academic Research Fund, under Award No. R-265-000-364-133.

## Author contributions

P.J. designed and carried out the experiments, conducted the data analysis and wrote the manuscript with the guidance and assistance of Y.K.K.

## Additional information

**Supplementary information** accompanies this paper at

http://www.nature.com/scientificreports

**Competing financial interests:** The authors declare no competing financial interests.

**Figures:**

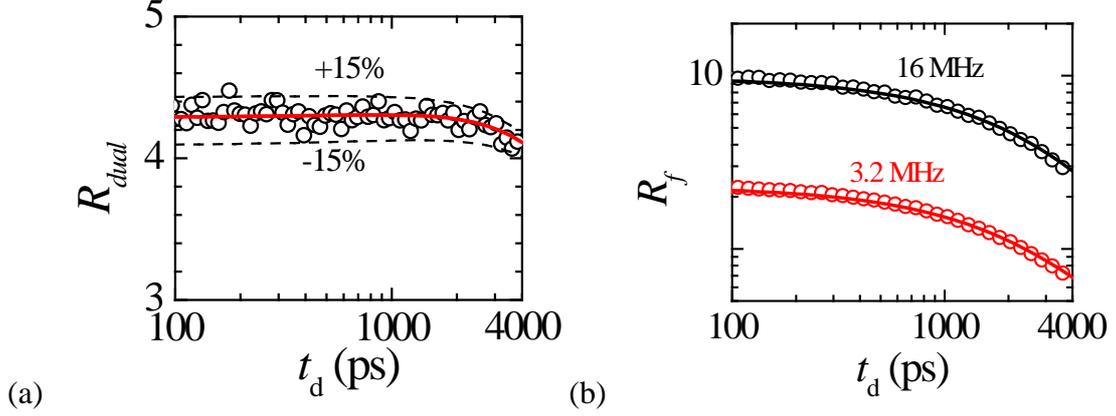

(a)  (b)

**Figure 1** | TDTR measurements (open circles) of a 1.02 µm thick Si film measured at $f_h$=16 MHz and $f_0$=3.2 MHz, plotted as (a) $R_{dual} = R_{f_h} / R_{f_0}$; and (b) $R_f = -V_{in}/V_{out}$ at each frequency $f$ as labeled. The solid lines are calculations of the thermal model using the $\Lambda_{cross}$ value fitted from the $R_{dual}$. The dashed lines are calculations of the thermal model using ±15% of the best-fit $\Lambda_{cross}$ value.

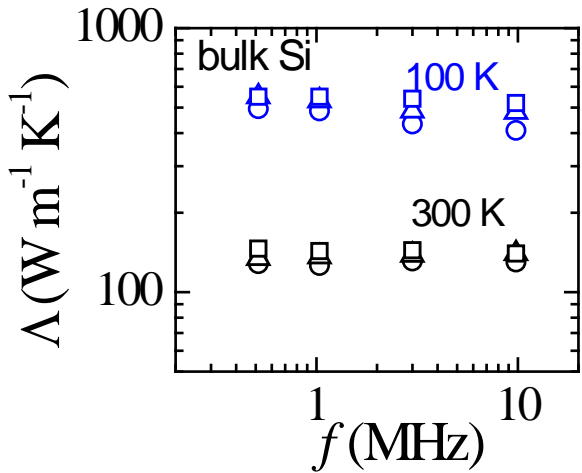

**Figure 2** | The apparent thermal conductivity $\Lambda$ of bulk Si measured as a function of modulation frequency $f$, using $1/e^2$ radii of laser beams of $w_0$=27 µm (squares), 11 µm (circles) and 5.5 µm (triangles) at 100 K and 300 K. The apparent thermal conductivity is independent of $f$ when the laser spot size is sufficiently large ($w_0$=27 µm).

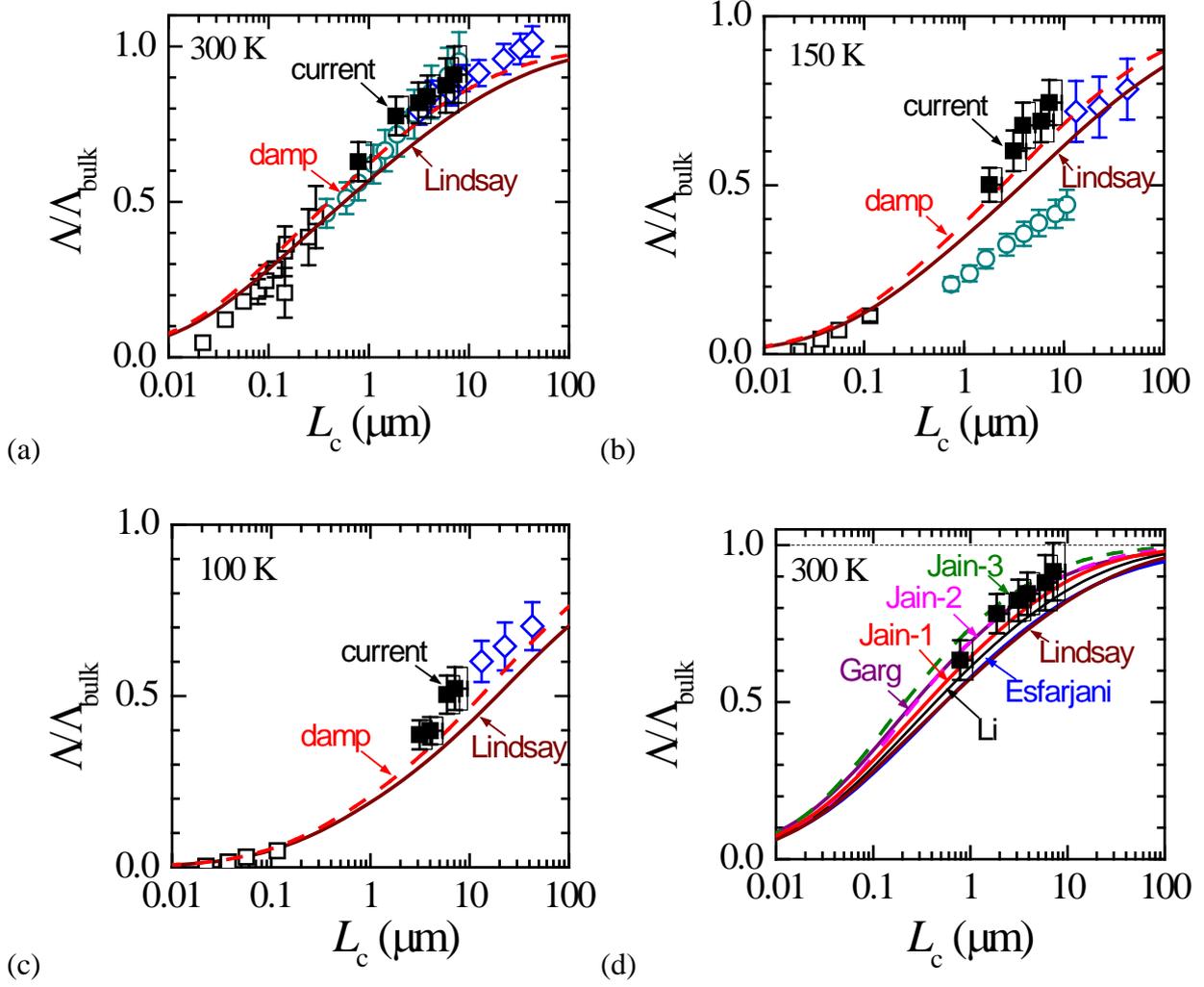

**Figure 3** | (a-c) $\Lambda_{cross}$ measurements of Si films (solid squares, current work) at (a) 300 K, (b) 150 K and (c) 100 K, compared with the thermal conductivity of Si nanowires[42-44] (open squares) and the apparent thermal conductivity of bulk Si measured by BB-FDTR[25] (open circles) and spot size dependent TDTR[26, 45] (open diamonds), plotted as a function of characteristic lengths $L_c$. All measurements are normalized by the thermal conductivity of bulk Si[41] of $\Lambda_{bulk}$=143 W m$^{-1}$ K$^{-1}$, 399 W m$^{-1}$ K$^{-1}$ and 853 W m$^{-1}$ K$^{-1}$ at $T$=300 K, 150 K, and 100 K, respectively. $L_c$ are $3h_f/4$ for $\Lambda_{cross}$ measurements,[46] the diameter[23] for the thermal conductivity of nanowires, the thermal penetration depth[25] $d_p$ for the BB-FDTR measurements, and the root-mean-square average of the pump and probe $1/e^2$ diameters[29] for the spot size dependent TDTR measurements, respectively. The shades in the figures represent the

uncertainty in the $\Lambda_{\text{cross}}$ and $L_c$. The solid lines are predictions derived from the first-principles calculations of natural Si by Lindsay et al.[11, 12] The dashed lines are calculations of our RTA model, taking into account the macroscopic damping of low-energy phonons, see the main text for the details of our RTA model. (d) Comparison of the $\Lambda_{\text{cross}}$ measurements in (a) with predictions derived from the first-principles calculations of pure Si by Lindsay et al.[11] (solid brown line), Li et al.[47] (solid black line), Esfarjani et al.[4] (solid blue line), Garg et al.[13] (solid purple line), and Jain and McGaughey[14] using different pseudopotential and exchange correlation schemes. (Jain-1 (solid red line) refers to calculations using a norm-conserving pseudopotential and the local density approximation (LDA); Jain-2 and Jain-3 (dashed pink and green lines) refer to calculations using the generalized gradient approximation (GGA) developed by different researchers, see calculations labeled "PW91" and "BYLP" in ref. [14] for more details). The first principles calculations are normalized by the bulk thermal conductivity calculated using the respective scheme.

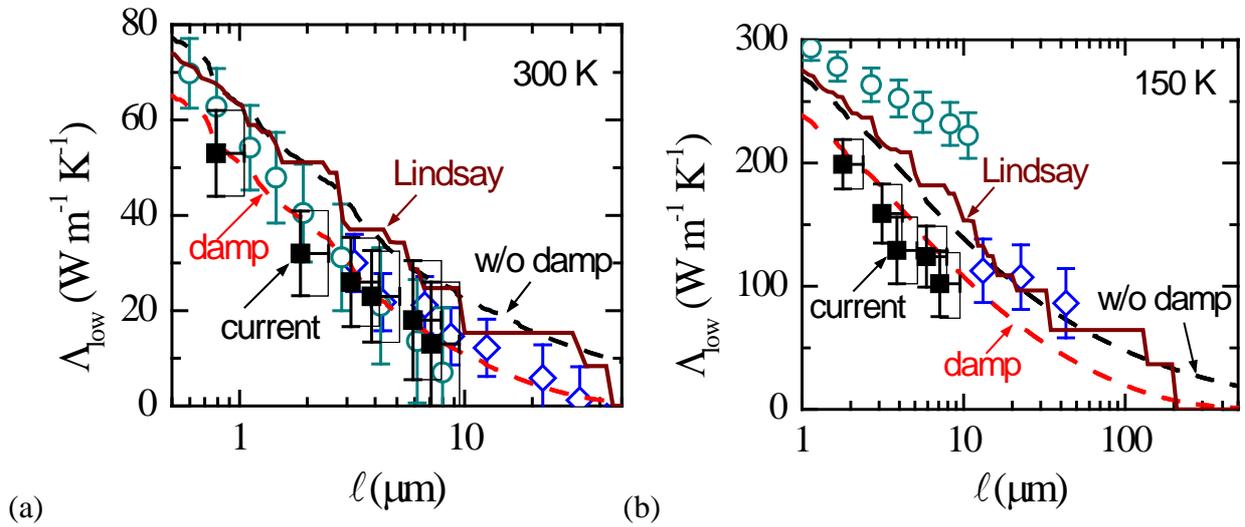

(a)  (b)

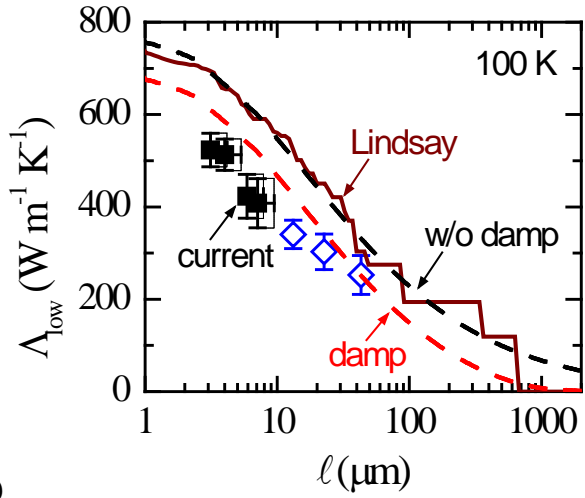

(c)

**Figure 4** | Cumulative thermal conductivity for low-energy phonons, $\Lambda_{low}$, in natural Si derived from $\Lambda_{cross}$ measurements of Si films (solid squares, current work) at (a) 300 K, (b) 150 K and (c) 100 K, compared with $\Lambda_{low}$ derived from the measurements of BB-FDTR[25] (open circles) and spot size dependent TDTR[26, 45] (open diamonds), as a function of phonon mean-free-path. The brown solid lines are the first principles calculations of natural Si by Lindsay *et al*. The dashed red and black lines are calculations of our RTA model, with and without incorporating the macroscopic damping, as labeled.

1. Ziman JM. *Electrons and Phonons: The Theory of Transport Phenomena in Solids*. Clarendon Press (1960).

2. Cahill DG, *et al.* Nanoscale thermal transport. II. 2003–2012. *Appl. Phys. Rev.* **1**, 011305 (2014).

3. Broido DA, Malorny M, Birner G, Mingo N, Stewart DA. Intrinsic lattice thermal conductivity of semiconductors from first principles. *Appl. Phys. Lett.* **91**, 231922 (2007).

4. Esfarjani K, Chen G, Stokes HT. Heat transport in silicon from first-principles calculations. *Phys. Rev. B* **84**, (2011).

5. Garg J, Bonini N, Kozinsky B, Marzari N. Role of Disorder and Anharmonicity in the Thermal Conductivity of Silicon-Germanium Alloys: A First-Principles Study. *Phys. Rev. Lett.* **106**, (2011).

6. Ward A, Broido D, Stewart D, Deinzer G. Ab initio theory of the lattice thermal conductivity in diamond. *Phys. Rev. B* **80**, (2009).

7. Luo T, Garg J, Shiomi J, Esfarjani K, Chen G. Gallium arsenide thermal conductivity and optical phonon relaxation times from first-principles calculations. *EPL (Europhysics Letters)* **101**, 16001 (2013).

8. Tian Z, Garg J, Esfarjani K, Shiga T, Shiomi J, Chen G. Phonon conduction in PbSe, PbTe, and PbTe$_{1-x}$Se$_x$ from first-principles calculations. *Phys. Rev. B* **85**, (2012).

9. Li W, Lindsay L, Broido DA, Stewart DA, Mingo N. Thermal conductivity of bulk and nanowire Mg$_2$Si$_x$Sn$_{1-x}$ alloys from first principles. *Phys. Rev. B* **86**, (2012).

10. Lindsay L, Broido DA, Reinecke TL. First-Principles Determination of Ultrahigh Thermal Conductivity of Boron Arsenide: A Competitor for Diamond? *Phys. Rev. Lett.* **111**, (2013).

11. Lindsay L, Broido D, Reinecke T. *Ab initio* thermal transport in compound semiconductors. *Phys. Rev. B* **87**, (2013).

12. Lindsay L, Broido D. Private communication. *(Private communication)*, (2014).

13. Garg J, Bonini N, Marzari N. First-Principles Determination of Phonon Lifetimes, Mean Free Paths, and Thermal Conductivities in Crystalline Materials: Pure Silicon and Germanium. In: *Length-Scale Dependent Phonon Interactions* (ed^(eds Shindé SL, Srivastava GP). Springer New York (2014).

14. Jain A, McGaughey AJH. Effect of exchange–correlation on first-principles-driven lattice thermal conductivity predictions of crystalline silicon. *Computational Materials Science* **110**, 115-120 (2015).

15. Maris HJ. 6 - Interaction of Sound Waves with Thermal Phonons in Dielectric Crystals. In: *Physical Acoustics* (ed^(eds Warren P M, R.N T). Academic Press (1971).

16. Daly BC, Kang K, Y. W, Cahill DG. Picosecond ultrasonic measurements of attenuation of longitudinal acoustic phonons in silicon. *Phys. Rev. B* **80**, (2009).

17. Duquesne JY, Perrin B. Ultrasonic attenuation in a quasicrystal studied by picosecond acoustics as a function of temperature and frequency. *Phys. Rev. B* **68**, (2003).

18. Hao HY, Maris H. Dispersion of the long-wavelength phonons in Ge, Si, GaAs, quartz, and sapphire. *Phys. Rev. B* **63**, (2001).

19. Maznev AA. Onset of size effect in lattice thermal conductivity of thin films. *J. Appl. Phys.* **113**, 113511 (2013).

20. Kim JH, Feldman A, Novotny D. Application of the three omega thermal conductivity measurement method. *J. Appl. Phys.* **86**, (1999).

21. Asheghi M, Touzelbaev MN, Goodson KE, Leung YK, Wong SS. Temperature-Dependent Thermal Conductivity of Single-Crystal Silicon Layers in SOI Substrates. *J. Heat Transfer* **120**, 30 (1998).

22. Cuffe J, *et al.* Reconstructing phonon mean-free-path contributions to thermal conductivity using nanoscale membranes. *Phys. Rev. B* **91**, (2015).

23. Sondheimer EH. The Mean Free Path of Electrons in Metals. *Adv. Phys.* **1**, (1952).

24. Koh Y, Cahill D. Frequency dependence of the thermal conductivity of semiconductor alloys. *Phys. Rev. B* **76**, (2007).

25. Regner KT, Sellan DP, Su Z, Amon CH, McGaughey AJ, Malen JA. Broadband phonon mean free path contributions to thermal conductivity measured using frequency domain thermoreflectance. *Nat. Commun.* **4**, 1640 (2013).

26. Minnich AJ, *et al.* Thermal conductivity spectroscopy technique to measure phonon mean free paths. *Phys. Rev. Lett.* **107**, (2011).

27. Johnson JA, *et al.* Direct Measurement of Room-Temperature Nondiffusive Thermal Transport Over Micron Distances in a Silicon Membrane. *Phys. Rev. Lett.* **110**, (2013).

28. Siemens ME, *et al.* Quasi-ballistic thermal transport from nanoscale interfaces observed using ultrafast coherent soft X-ray beams. *Nat Mater* **9**, 26-30 (2010).

29. Wilson RB, Cahill DG. Anisotropic failure of Fourier theory in time-domain thermoreflectance experiments. *Nat. Commun.* **5**, 5075 (2014).

30. Koh YK, Cahill DG, Sun B. Nonlocal theory for heat transport at high frequencies. *Phys. Rev. B* **90**, (2014).

31. Yang F, Dames C. Mean free path spectra as a tool to understand thermal conductivity in bulk and nanostructures. *Phys. Rev. B* **87**, (2013).

32. Celler GK, Cristoloveanu S. Frontiers of silicon-on-insulator. *J. Appl. Phys.* **93**, 4955 (2003).

33. Koh YK, Vineis CJ, Calawa SD, Walsh MP, Cahill DG. Lattice thermal conductivity of nanostructured thermoelectric materials based on PbTe. *Appl. Phys. Lett.* **94**, 153101 (2009).

34. Hohensee GT, Hsieh WP, Losego MD, Cahill DG. Interpreting picosecond acoustics in the case of low interface stiffness. *Rev. Sci. Instrum.* **83**, 114902 (2012).

35. Strohmeier BR. An ESCA method for determining the oxide thickness on aluminum alloys. *Surf. Interface Anal.* **15**, 51-56 (1990).

36. See online supporting material.

37. A. C, Paddock, Eesley GL. Transient thermoreflectance from thin metal films. *J. Appl. Phys.* **60**, 285 (1986).

38. Cahill DG. Analysis of heat flow in layered structures for time-domain thermoreflectance. *Rev. Sci. Instrum.* **75**, 5119 (2004).

39. Jiang P, Koh YK. Accurate measurements of cross-plane thermal conductivity of thin Films by dual-Frequency time-domain thermoreflectance (TDTR). *Rev. Sci. Instrum.* **Submitted**.

40. Schmidt AJ, Cheaito R, Chiesa M. A frequency-domain thermoreflectance method for the characterization of thermal properties. *Rev. Sci. Instrum.* **80**, 094901 (2009).

41. Kremer RK, *et al.* Thermal conductivity of isotopically enriched $^{28}$Si: revisited. *Solid State Commun.* **131**, 499-503 (2004).

42. Li D, Wu Y, Kim P, Shi L, Yang P, Majumdar A. Thermal conductivity of individual silicon nanowires. *Appl. Phys. Lett.* **83**, 2934 (2003).

43. Doerk GS, Carraro C, Maboudian R. Single nanowire thermal conductivity measurements by Raman thermography. *ACS nano* **4**, 4908-4914 (2010).

44. Liu D, Xie R, Yang N, Li B, Thong JT. Profiling nanowire thermal resistance with a spatial resolution of nanometers. *Nano Lett.* **14**, 806-812 (2014).

45. Ding D, Chen X, Minnich AJ. Radial quasiballistic transport in time-domain thermoreflectance studied using Monte Carlo simulations. *Appl. Phys. Lett.* **104**, 143104 (2014).

46. Majumdar A. Microscale heat conduction in dielectric thin films. *J. Heat Transfer* **115**, 7 (1993).

47. Li W, Mingo N, Lindsay L, Broido DA, Stewart DA, Katcho NA. Thermal conductivity of diamond nanowires from first principles. *Phys. Rev. B* **85**, (2012).

48. Minnich AJ. Determining Phonon Mean Free Paths from Observations of Quasiballistic Thermal Transport. *Phys. Rev. Lett.* **109**, (2012).

49. Hu Y, Zeng L, Minnich AJ, Dresselhaus MS, Chen G. Spectral mapping of thermal conductivity through nanoscale ballistic transport. *Nat Nano* **10**, 701-706 (2015).

50. Zebarjadi M, Esfarjani K, Dresselhaus M, Ren Z, Chen G. Perspectives on thermoelectrics: from fundamentals to device applications. *Energy & Environmental Science* **5**, 5147-5162 (2012).

51. Koh YK, Cao Y, Cahill DG, Jena D. Heat-Transport Mechanisms in Superlattices. *Adv. Funct. Mater.* **19**, 610-615 (2009).

52. Kor S, Mishra P, Tripathi N. Ultrasonic attenuation in pure and doped n-type silicon. *Phys. Rev. B* **10**, 775-778 (1974).

53. Weber W. Adiabatic bond charge model for the phonons in diamond, Si, Ge, and α-Sn. *Phys. Rev. B* **15**, 4789-4803 (1977).

## Supplementary information:

# The role of low-energy phonons with mean-free-paths >0.8 μm in heat conduction in silicon

## Puqing Jiang and Yee Kan Koh[*]

Department of Mechanical Engineering, National University of Singapore, Singapore 117576

[*]corresponding author: mpekyk@nus.edu.sg

## I. Uncertainty analysis of $\Lambda_{\text{cross}}$ measured by dual-frequency TDTR approach

In the dual-frequency approach, we fit the ratio of TDTR signals at the two frequencies $R_{dual} = R_{f_1}/R_{f_0}$, instead of $R_f$. The sensitivity of $R_{dual}$ is the difference of the sensitivities of TDTR signals at the two frequencies:

$$S_\alpha^{dual} = \frac{\partial \ln\left(R_{f_1}/R_{f_0}\right)}{\partial \ln \alpha} = \frac{\partial \ln R_{f_1}}{\partial \ln \alpha} - \frac{\partial \ln R_{f_0}}{\partial \ln \alpha} = \left.S_\alpha^{\text{TDTR}}\right|_{f_1} - \left.S_\alpha^{\text{TDTR}}\right|_{f_2} \qquad (S1)$$

In this way, we maintain the sensitivity to $\Lambda_{\text{cross}}$, but significantly reduce the sensitivities to the other parameters, see Fig. S1(a-b).

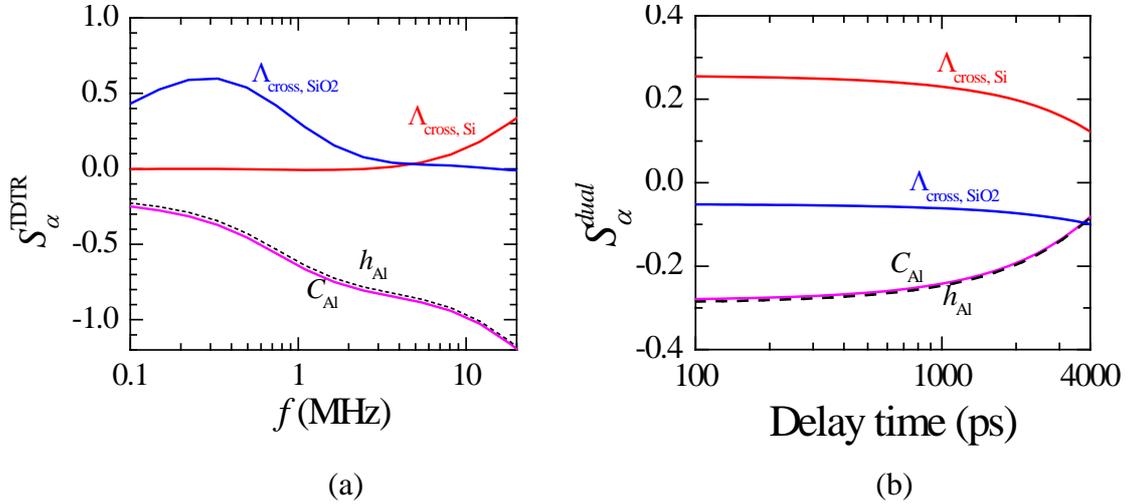

(a)  (b)

**Figure S1** | (a) Sensitivity of TDTR $R_f$ signal to $\Lambda_{\text{cross}}$, $h_{\text{Al}}$ and $C_{\text{Al}}$, as a function of modulation frequency. The sample is a 1 μm thick Si film on a 380 nm thick SiO$_2$ film on Si substrate, coated with a 100 nm thick Al film. The delay time is fixed at 100 ps and the laser spot radius is 27 μm. (b) Sensitivity of $R_{dual} = R_{f_1}/R_{f_0}$, with $f_1$=16 MHz and $f_0$=3.2 MHz, to parameters $\Lambda_{\text{cross}}$, $h_{\text{Al}}$ and $C_{\text{Al}}$.

The uncertainty of $\Lambda_{\text{cross}}$ of the Si film is calculated as:

$$\eta_{\Lambda_{\text{cross}}} = \sqrt{\sum\left(\frac{S_{\alpha}^{dual}}{S_{\Lambda_{\text{cross}}}^{dual}}\eta_{\alpha}\right)^2 + \left(\frac{S_{\phi_1}^{dual}}{S_{\Lambda_{\text{cross}}}^{dual}}\delta\phi_1\right)^2 + \left(\frac{S_{\phi_2}^{dual}}{S_{\Lambda_{\text{cross}}}^{dual}}\delta\phi_2\right)^2} \qquad (S2)$$

Input parameters $\alpha$ in the thermal model include the laser spot size, and the thickness, heat capacity, cross-plane and in-plane thermal conductivity of each layer of the sample. In the analysis, $\Lambda_{\text{cross}}$ of the Si thin film and the Al/Si interface conductance $G$ are treated as freely adjustable parameters. For the Al/Si interface conductance $G$, we consistently get the values of 330, 230 and 160 MW m$^{-2}$ K$^{-1}$ at temperatures 300 K, 150 K and 100 K respectively, independent of Si film thickness. TDTR is not sensitive to the Si/SiO$_2$ interface conductance $G_2$; we have varied the $G_2$ value over a wide range from 50 MW m$^{-2}$ K$^{-1}$ to 100 GW m$^{-2}$ K$^{-1}$, and find it affects $\Lambda_{\text{cross}}$ by <2%.

In our TDTR measurements, we estimate the uncertainty of each input parameter as follows: thermal conductivities $\Lambda_{\text{Al}}$ as 15%, $\Lambda_{\text{SiO2}}$ as 5%; heat capacities $C_{\text{Al}}$ as 5%, $C_{\text{Si}}$ as 5%, $C_{\text{SiO2}}$ as 5%; thicknesses $h_{\text{Al}}$ as 6.5%, $h_{\text{Si}}$ as 4%, $h_{\text{SiO2}}$ as 4%; interface conductance $G$ as 25%; and spot size $w$ as 8%. Based on the uncertainties and sensitivities of all the parameters, we estimate the uncertainties of our $\Lambda_{\text{cross}}$ measurements as ~10%.

## II. Estimate thickness of native Al oxide by XPS

Figure S2 is the XPS spectrum of our sample of a thermally evaporated Al film exposed to the typical laboratory environment of Singapore for two weeks. Based on the method of Strohmeier, (see Eq. (2) in Ref. [1]), we estimate the oxide film thickness as ~37.4 Å for this specific sample.

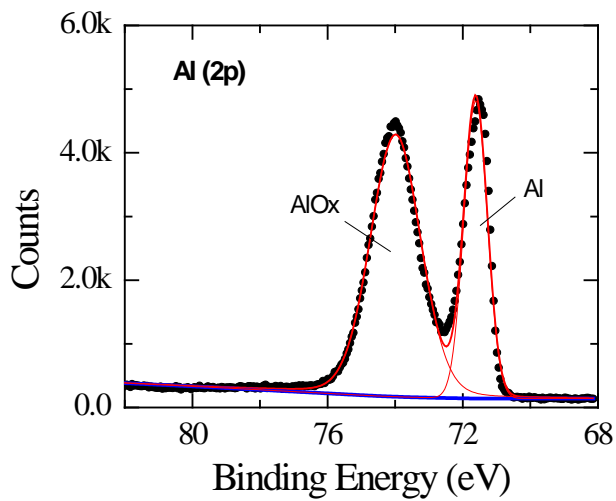

**Figure S2** | XPS spectrum of a thermally evaporated Al film with a native oxide layer. The Al 2p peak envelop shows the respective metal and metal oxide components, from which the $AlO_x$ layer thickness can be estimated.

## III. Estimate effect of macroscopic damping using a relaxation time approximation (RTA) model

The first-principles calculations do not take into account the macroscopic damping of low-energy phonons. To estimate the effect of macroscopic damping, we use a relaxation time approximation (RTA) model that reproduces the cumulative thermal conductivity of natural Si by first-principles calculations of Lindsay et al.[2, 3]

In the RTA model, the thermal conductivity is calculated as:

$$\Lambda = \frac{1}{3}\sum_j \int_0^\infty \frac{\hbar^2\omega^2}{k_B T^2} n_0(n_0+1)\tau_c v^2 D(\omega)d\omega \qquad (S3)$$

where $j$ stands for polarization, $\hbar$ is the reduced Planck constant, $k_B$ is the Boltzmann constant, $n_0$ is the Planck distribution function, $\tau_c$ is the combined relaxation time of relaxation times for N-process $\tau_N$, U-process $\tau_U$ and impurity scattering $\tau_I$, determined by Matthiessen's rule as $\tau_c^{-1} = \tau_N^{-1} + \tau_U^{-1} + \tau_I^{-1}$, $v$ is the group velocity, $D(\omega)$ is the density of states, $\omega$ is the phonon angular frequency. The ingredient $v^2 D(\omega)$ of Si in the integral in Eq. (S3) is calculated utilizing a full phonon dispersion from the adiabatic bond charge model[4] and the Gilat-Raubenheimer method[5] for integration over the k-space.

We extract the relaxation times of the normal and Umklapp processes in Si from the zeroth-order solution of the first-principles calculations by Lindsay et al.[2, 3] For high-energy phonons with frequency $\omega/2\pi > 3$ THz, we estimate the relaxation times from the harmonic mean of the first-principles calculations. For the low-energy phonons with $\omega/2\pi < 3$ THz, we assume a $\omega^2$ dependence for the N-process and a $\omega^3$ dependence for the U-process scattering rates[6] and extrapolate relaxation times for the low-energy phonons from the limited first-principles calculations data, see Figure S3.

We calculate the relaxation time for impurity scattering from[7]

$$\tau_I(\omega)^{-1} = \frac{\pi V}{6}\Gamma \omega^2 D(\omega) \qquad (S4)$$

where $V$ is the volume per atom, $\Gamma$ is the dimensionless mass-defect scattering strength and can be calculated[8] as $\Gamma = \sum_i c_i (1 - m_i/\bar{m})^2$, with $c_i$ the fractional concentration of the $i$th species, $m_i$ the atomic mass the $i$th species, and $\bar{m}$ the average atomic mass.

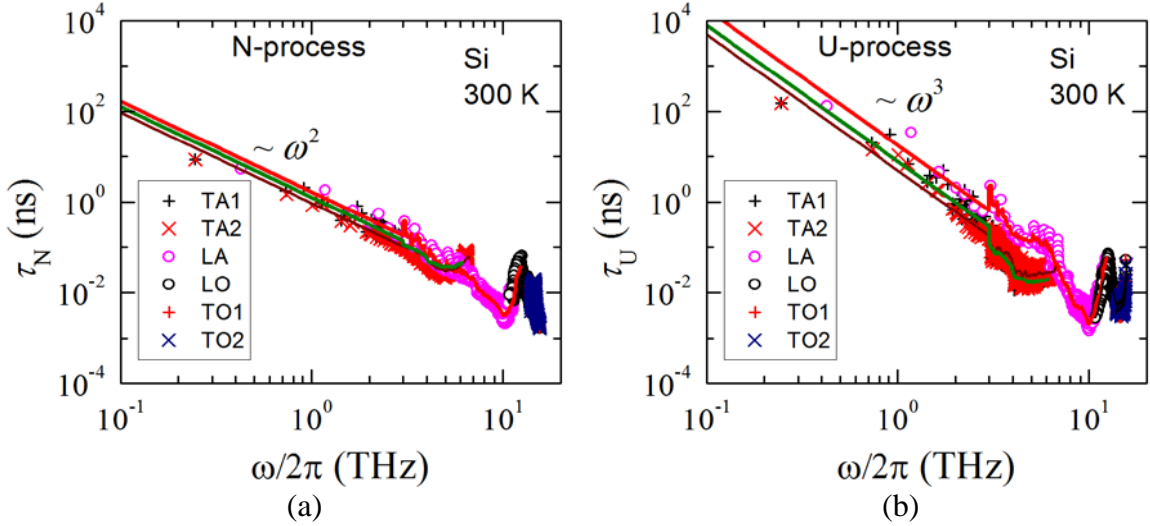

**Figure S3** | Relaxation times of Si for (a) N-process and (b) U-process, from the zeroth-order solution of first-principles calculation at 300 K. The symbols are the direct outputs from first-principles calculations, and the curves are extracted relaxation times used in our RTA model.

To estimate the effect of macroscopic damping on low-frequency phonons, we follow Maznev[9] to add an additional scattering term to the total relaxation time. We employ Matthiessen's rule and derive $\tau_c^{-1} = \tau_N^{-1} + \tau_U^{-1} + \tau_I^{-1} + \tau_A^{-1}$, where $\tau_A$ is the relaxation time contribution due to the macroscopic damping

$$\tau_A = \tau_{inf}\left(1 + \frac{1}{\tau_{th}^2 \omega^2}\right) \tag{S5}$$

The parameters $\tau_{inf}$ and $\tau_{th}$ are the fitting parameters derived by comparing the calculation of $\tau_c$ to the available experimental data.[10-15] The derived values of $\tau_{inf}$ and $\tau_{th}$ at different temperatures are summarized in Table S1.

The cumulative thermal conductivity of natural Si calculated by the RTA model without macroscopic damping compare very well with the first-principles results at different temperatures, as shown in Fig. S4 as the solid lines. The effect of macroscopic damping estimated on the basis of the RTA model is shown as the dashed lines in Fig. S4.

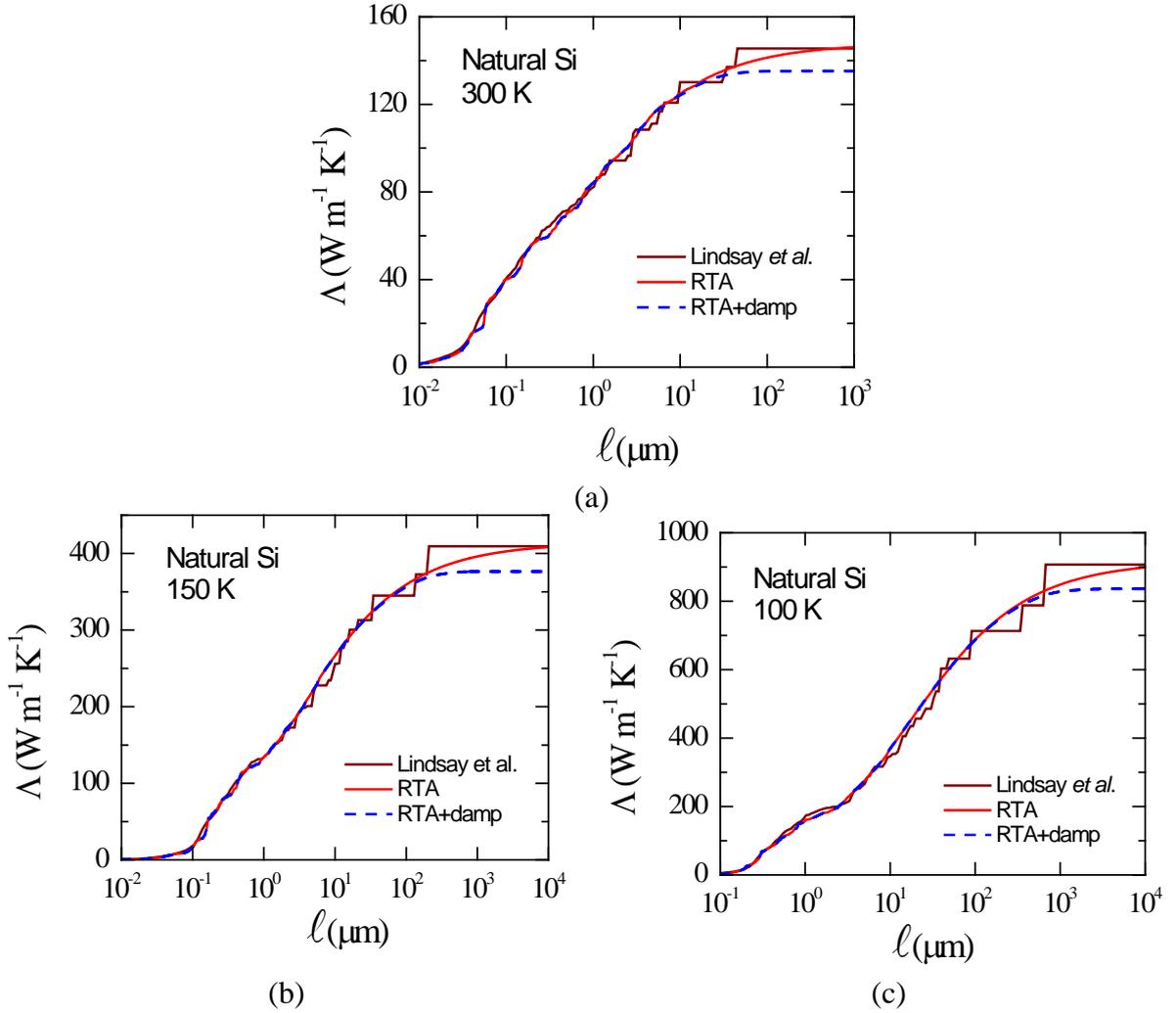

**Figure S4** | Cumulative thermal conductivity of natural Si at 300 K, 150 K and 100 K calculated by RTA model with and without incorporating macroscopic damping, compared with the first-principles results by Lindsay *et al.*

**Table S1:** Parameters for the relaxation times of macroscopic damping

|  | $\tau_{\text{inf}}$ (ns) |  | $\tau_{\text{th}}$ (ps) |
|---|---|---|---|
|  | LA | TA |  |
| 100 K | 140 | 700 | 50 |
| 150 K | 30 | 150 | 30 |
| 300 K | 5 | 25 | 14 |

## IV. Approximation of $\Lambda_{low}$ from $\Lambda_{bulk}-\Lambda_{cross}$

In the main text, we approximate the cumulative thermal conductivity of low-energy phonons $\Lambda_{low}$ in bulk Si as $\Lambda_{low}=\Lambda_{bulk}-\Lambda_{cross}$ from our cross-plane thermal conductivity measurements. To evaluate the accuracy of this approximation, we calculate $\Lambda_{bulk}-\Lambda_{cross}$ from $\Lambda_{cross}$ of Si thin films derived from the first-principles calculations of the accumulation function of bulk Si (see the main text for how the conversion is performed), and compare the calculations to the $\Lambda_{low}$ of bulk Si derived directly from the first-principles calculations using the definition $\Lambda_{low}=\int_{\ell}^{\infty}\Lambda_{\ell_{bulk}}d\ell_{bulk}$, see Fig. S5. We observe that within the mean-free-path of $0.1<\ell<10$ μm, $\Lambda_{bulk}-\Lambda_{cross}$ is an acceptable approximation for $\Lambda_{low}$, see Fig. S5.

To understand the observation, we define $\Delta=\Lambda_{low}-\left(\Lambda_{bulk}-\Lambda_{cross}\right)$ as the error for using the expression $\Lambda_{bulk}-\Lambda_{cross}$. To the first-order approximation,

$$\Delta=\int_{L_c}^{\infty}\Lambda_{\ell_{bulk}}\frac{L_c}{\ell_{bulk}}d\ell_{bulk}-\int_{0}^{L_c}\Lambda_{\ell_{bulk}}\frac{\ell_{bulk}}{L_c}d\ell_{bulk} \tag{S6}$$

where $\ell_{bulk}$ is the mean-free-path of phonons in bulk Si, $L_c=3h_f/4$ is the characteristic length for boundary scattering in Si thin films, and $\Lambda_{\ell_{bulk}}=-\sum_{j}\frac{1}{3}Cv\ell_{bulk}\frac{d\omega}{d\ell_{bulk}}$ is the bulk thermal conductivity per mean-free-path,[16] with an SI unit of W m$^{-2}$ K$^{-1}$. We have utilized Matthiessen's rule to relate $\ell_f$ of thin film and $\ell_{bulk}$ of bulk Si, $\ell_f^{-1}=\ell_{bulk}^{-1}+L_c^{-1}$, when deriving the error $\Delta$.

The first term of Eq. S6 represents the error due to underestimation of the contribution by phonons with $\ell>L_c$ using $\Lambda_{bulk}-\Lambda_{cross}$, because of the additional boundary scattering at $L_c$ in $\Lambda_{cross}$. The second term, on the other hand, represents the residues of contribution by phonons with $\ell<L_c$ in $\Lambda_{bulk}-\Lambda_{cross}$, compared to zero contribution from these phonons in $\Lambda_{low}$. Note that the first and the second terms always have opposite signs. Within $0.1<\ell<10$ μm, the amplitude of both terms are comparable and thus leaving $\Delta$ close to zero. As $\ell\to\infty$, the first term is diminishing and, as a result, $\Lambda_{bulk}-\Lambda_{cross}$ becomes larger than $\Lambda_{low}$. This is observed in Fig. S5 when $\ell$ >1 μm. Thus, within the range of our experiments, the expression $\Lambda_{bulk}-\Lambda_{cross}$ is an over-approximation of $\Lambda_{low}$ and set an upper limit for $\Lambda_{low}$.

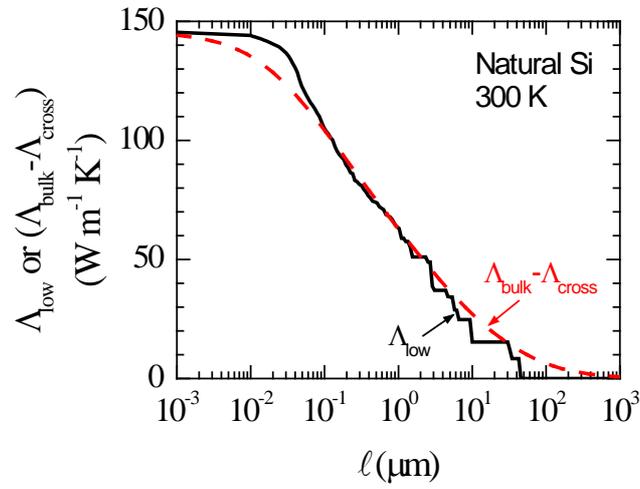

**Figure S5** | $\Lambda_{low}$ of natural Si at 300 K as a function of phonon mean-free-paths calculated by the first-principles approach of Lindsay *et al*., compared with the approximation from measurements of ($\Lambda_{bulk}$-$\Lambda_{cross}$) plotted as a function of $L_c$.

## V. Comparison of the accumulation functions of natural and isotropically pure Si

For some first-principles calculations, only the accumulation functions of isotopically pure Si are available in literature. We demonstrate in Fig. S6 that accumulation functions of natural Si and pure Si from the first-principles calculations of Lindsay *et al*.[2, 3] are almost identical above 100 K. Thus it is valid to compare the measurements of natural Si with calculations of pure Si in terms of thermal conductivity ratio or accumulation function.

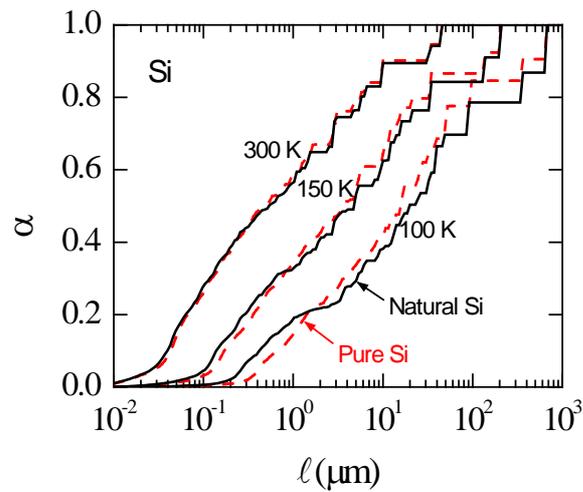

**Figure S6** | Comparison of accumulation functions of natural Si vs. pure Si calculated by Lindsay *et al*.[2]

## VI. Comparison of different first-principles calculations on Si

Here we compile the thermal conductivity accumulation functions of pure Si at 300 K calculated by different research groups via first-principles approach. It shows that except the calculation by Jain and McGaughey[17] using the generalized gradient approximation (GGA) developed by Becke[18] and Lee et al.[19] (labeled "BLYP" in Ref. [17] and referred as Jain-2 in this plot), all the first-principles approaches result in similar thermal conductivity accumulation, up to ≈5 µm. For the long mean-free-path range, however, the calculations by different groups differ from each other, suggesting the difficulties in the correct calculation of low-energy phonons.

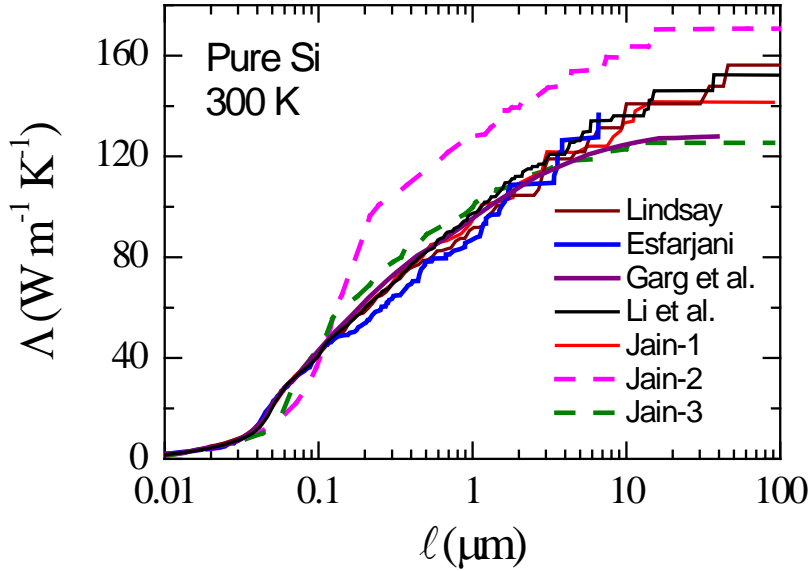

**Figure S7** | Accumulation functions of pure Si at 300 K from first-principles calculations of different research groups by Lindsay et al.,[2, 3] Li et al.,[20] Esfarjani et al.,[6, 16] Garg et al.,[21] and Jain and McGaughey[17] using different schemes (Jain-1 refers to calculations using a norm-conserving pseudopotential and the local density approximation (LDA); Jain-2 and Jain-3 refer to calculations using the generalized gradient approximation (GGA) developed by different researchers, see calculations labeled "PW91" and "BYLP" in ref. [17] for more details).

**Supplementary References:**

1. Strohmeier BR. An ESCA method for determining the oxide thickness on aluminum alloys. *Surf. Interface Anal.* **15**, 51-56 (1990).

2. Lindsay L, Broido D, Reinecke T. *Ab initio* thermal transport in compound semiconductors. *Phys. Rev. B* **87**, (2013).

3. Lindsay L, Broido D. Private communication. *(Private communication)*, (2014).

4. Weber W. Adiabatic bond charge model for the phonons in diamond, Si, Ge, and α-Sn. *Phys. Rev. B* **15**, 4789-4803 (1977).

5. Gilat G, Raubenheimer L. Accurate Numerical Method for Calculating Frequency-Distribution Functions in Solids. *Phys. Rev.* **144**, 390-395 (1966).

6. Esfarjani K, Chen G, Stokes HT. Heat transport in silicon from first-principles calculations. *Phys. Rev. B* **84**, (2011).

7. Tamura S-i. Isotope scattering of dispersive phonons in Ge. *Phys. Rev. B* **27**, 858-866 (1983).

8. Cahill D, Watanabe F, Rockett A, Vining C. Thermal conductivity of epitaxial layers of dilute SiGe alloys. *Phys. Rev. B* **71**, (2005).

9. Maznev AA. Onset of size effect in lattice thermal conductivity of thin films. *J. Appl. Phys.* **113**, 113511 (2013).

10. Daly BC, Kang K, Y. W, Cahill DG. Picosecond ultrasonic measurements of attenuation of longitudinal acoustic phonons in silicon. *Phys. Rev. B* **80**, (2009).

11. Mason WP, Bateman TB. Ultrasonic-Wave Propagation in Pure Silicon and Germanium. *J. Acoust. Soc. Am.* **36**, 644-652 (1964).

12. Pomerantz M. Temperature Dependence of Microwave Phonon Attenuation. *Phys. Rev.* **139**, A501-A506 (1965).

13. Keller KR. Ultrasonic Attenuation in Ge-Si Alloys. *J. Appl. Phys.* **38**, 3777 (1967).

14. Hao HY, Maris H. Dispersion of the long-wavelength phonons in Ge, Si, GaAs, quartz, and sapphire. *Phys. Rev. B* **63**, (2001).

15. Duquesne JY, Perrin B. Ultrasonic attenuation in a quasicrystal studied by picosecond acoustics as a function of temperature and frequency. *Phys. Rev. B* **68**, (2003).

16. Yang F, Dames C. Mean free path spectra as a tool to understand thermal conductivity in bulk and nanostructures. *Phys. Rev. B* **87**, (2013).

17. Jain A, McGaughey AJH. Effect of exchange–correlation on first-principles-driven lattice thermal conductivity predictions of crystalline silicon. *Computational Materials Science* **110**, 115-120 (2015).

18. Becke AD. Density-functional exchange-energy approximation with correct asymptotic behavior. *Phys. Rev. A* **38**, 3098-3100 (1988).

19. Lee C, Yang W, Parr RG. Development of the Colle-Salvetti correlation-energy formula into a functional of the electron density. *Phys. Rev. B* **37**, 785-789 (1988).

20. Li W, Mingo N, Lindsay L, Broido DA, Stewart DA, Katcho NA. Thermal conductivity of diamond nanowires from first principles. *Phys. Rev. B* **85**, (2012).

21. Garg J, Bonini N, Marzari N. First-Principles Determination of Phonon Lifetimes, Mean Free Paths, and Thermal Conductivities in Crystalline Materials: Pure Silicon and Germanium. In: *Length-Scale Dependent Phonon Interactions* (ed^(eds Shindé SL, Srivastava GP). Springer New York (2014).